\documentclass[10pt,
    superscriptaddress,
    amsmath, amssymb,
    aps, pra,
    floatfix, nofootinbib, 
    twocolumn, showpacs,
    longbibliography
]{revtex4-1}

\usepackage{amsthm}
\usepackage{physics}
\usepackage{graphicx}% Include figure files
\usepackage{dcolumn}% Align table columns on decimal point
\usepackage{bm}% bold math
\usepackage[hyperindex,breaklinks]{hyperref}
\usepackage[T1]{fontenc}
\theoremstyle{definition}
\newtheorem{defn}{Definition}
\newtheorem{prop}{Proposition}

\newtheorem{lem}{Lemma}

\newtheorem{obs}{Observation}
\newtheorem{exa}{Example}

\hypersetup{
    breaklinks=true,   % splits links across lines
    colorlinks=true,   % displays links as colored text instead of blocks
    pdfusetitle=true,  % \title and \author values into pdf metadata
    citecolor = blue
    }

\begin{document}

\preprint{APS/123-QED}

\title{Entanglement witnesses with local partial ordering}% Force line breaks with \\

\author{Joshua Carlo A. Casapao}
 \email{joshuacarlo.casapao@oist.jp}
\affiliation{%
 Networked Quantum Devices Unit, Okinawa Institute of Science and Technology Graduate University, Okinawa, 904-0495 Japan\\
}%

\author{Eric A. Galapon}
 \email{eagalapon@up.edu.ph}
\affiliation{
 Theoretical Physics Group, National Institute of Physics,
University of the Philippines, Diliman, 1101 Philippines
}%

\date{\today}

\begin{abstract}
We investigate a class of entanglement witnesses where each witness is formulated as a difference of two product observables. These observables are decomposable into positive semidefinite local operators that obey a partial ordering rule defined over all their possible expectation values. We provide a framework to construct these entanglement witnesses along with some examples. We also discuss methods to improve them both linearly and nonlinearly. 

% \begin{description}
% Secondary publications and information retrieval purposes.
% You may use the \texttt{description} environment to structure your abstract;
% use the optional argument of the \verb+\item+ command to give the category of each item. 
% \end{description}
\end{abstract}

\maketitle

%--------------------%

\section{Introduction}

It is undeniable that entanglement enabled tremendous developments in quantum technologies, with it acting as a key ingredient in quantum information processes. Verifiable entanglement translates to quantum-computational speed-up \cite{Jozsa2003ontherole, datta2005entanglement}, deployable resources for secure quantum communication \cite{scarani2009security,wilde2013quantum}, and other emerging quantum applications \cite{Acin2018roadmap}. Hence, it is vital to develop our capability to detect or verify the presence of entanglement in an optimal and efficient manner, while keeping in mind the difficulty in experimental control \cite{Guhne2009entanglementdetection,Friis2018entanglement}. 

A reliable verification scheme is to perform a full quantum tomographic protocol \cite{hradil1997quantum,bisio2009optimal}, which essentially reconstructs a quantum state. Whenever entanglement is not immediately obvious from the reconstruction, the reconstruction can be additionally post-processed using known separability criteria, i.e., sufficient conditions for entanglement \cite{horodecki2009quantum,Guhne2009entanglementdetection}. Unfortunately, tomography suffers from a scalability problem, where the number of measurement settings needed increases exponentially with the size of the system. This renders the scheme cost-inefficient or even outright impossible for large complex systems. 

Alternatively, we can simply circumvent full tomography altogether and just directly implement one of the many scalar separability criteria \cite{horodecki2009quantum}, such as those based on the entropic and purity-based inequalities \cite{horodecki1996information,flores2016mixtures}. These schemes involve nonlinear functionals of density matrices that are typically difficult to extract in single-copy measurements, but become accessible with intricate collective measurements \cite{bovino2005direct,Islam2015measuring,brun2004measuring}. On the other hand, recent proposals show that polynomial functionals of a density matrix, as well as its partial transpose spectrum, can be efficiently estimated using a randomized measurement strategy, which opens up new methods to detect entanglement \cite{Huang2020predicting,elben2020mixed,neven2021symmetry,Elben2022randomized}. 

Entanglement witnesses (EWs) \cite{Chruscinski2018Entanglement, riccardi2020optimal,bae2020mirrored,Guhne2021Geometry,Wang2021Constructing,Frerot2021optimal,Greenwood2023Machine,Igloi2023Entanglement,rico2024entanglement} offer a practical solution to reveal entanglement without resorting to a full tomographic information, with many EW experiments already able to demonstrate verifying generated entanglement given the currently attainable technologies involving photonic systems and trapped ions \cite{Barbieri2003detection,bourennane2004experimental,Haffner2005scalable,Lu2007experimental,guo2018experimental}. EWs are directly measurable observables designed such that they have non-negative expectation values over the entire set of separable states, while a negative expectation value indicates a signature of entanglement \cite{Guhne2009entanglementdetection}. It is worth noting that a non-negative expectation value does not entail that the underlying state is separable, and that entanglement verification using the chosen EW is just undecidable. Crucially, every entangled state is guaranteed to be witnessed by some EW following the Hahn-Banach separation theorem \cite{Horodecki1996separability}. Although this ensures that EW theory is complete in terms of detecting power, constructing EWs for arbitrary states remains a difficult task unless some prior information about the produced states is available \cite{guhne2002detection,Friis2018entanglement}.

While EWs can in principle be measured in a single corresponding basis, this basis can only be described nonlocally and is therefore not a product basis that can be easily accessed in an experiment \cite{Friis2018entanglement}. As such, a typical approach in estimating EWs is to decompose them into experimentally accessible local observables (with number of these observables much smaller than what is needed for full tomography), and measure all these observables directly \cite{terhal2002detecting,guhne2002detection,guhne2003investigating}. An obvious catch to this approach is that there is an inevitable trade-off between minimizing an EW's measurement complexity and maximizing its noise tolerance, both of which determine the experimental achievability \cite{sciara2019universal}. On this account, several works have either constructed or customized EWs so that they can be compactly implemented in a few local measurements while retaining a modest robustness against noise \cite{toth2005entanglement,sciara2019universal,Zhou2019detecting}. Extending these observations to other complex and arbitrary-dimensional quantum systems remains an open problem. On the one hand, recent interests in using a randomized measurement strategy to efficiently estimate EWs may allow for flexibility in entanglement detection since we can simultaneously test multiple candidate EWs efficiently given that partial information about the state can be stored in a classical memory \cite{Huang2020predicting,rico2024entanglement}. \\

In this present contribution, we investigate a class of EWs, where each EW is formulated as a difference of two product observables. The only assumptions that we impose are that the corresponding local operators are positive semidefinite, and that they obey a partial ordering rule defined over all their possible expectation values. Essentially, we consider EWs that can be prescriptively constructed, while demanding them to be structurally simple by requiring that each EW reveal entanglement with effectively two local measurements. Our choice of definition is also partly motivated by the fact that the Bell-CHSH inequalities may also be written into witness-like inequalities comprising of measuring two product observables \cite{terhal2000bellineq,Hyllus2005relations}. \\

This work is structured as follows. In Section~\ref{sec:proposed_witness}, we introduce the collection of EWs that can be constructed by taking the difference of two operators that are both locally decomposable and both obey a local partial ordering rule. This rule will be further discussed in Section~\ref{sec:witness_construction}. We then provide a strategy to detect genuine multipartite entanglement with our proposed EWs in Section~\ref{sec:genuine_entanglement}. In Section~\ref{sec:witness_improvements}, we show that our proposed EWs can be improved both linearly and nonlinearly, while retaining the locally decomposable property. Finally, we conclude this work in Section~\ref{sec:conclusions}.

%--------------------%

\section{Proposed entanglement witnesses}
\label{sec:proposed_witness}

A quantum state $\rho$ defined over an $N$-composite Hilbert space $\mathcal{H}=\bigotimes_{j=1}^N\mathcal{H}_j$ belongs to the set of fully separable states $\mathcal{S}$ if can be written as a convex sum of product states,
\begin{equation}
    \rho = \sum_{q}\lambda_q\rho_{1,q}\otimes \rho_{2,q}\otimes\cdots\otimes\rho_{N,q}.
\end{equation}
Otherwise, we say that $\rho$ is entangled. We say that $\mathcal{W}$ is an EW to some entangled state $\rho$ if the following statements are true:
\begin{equation}
    \Tr(\mathcal{W}\sigma) \geq 0, \quad \forall \sigma\in \mathcal{S}; \qquad \Tr(\mathcal{W}\rho)<0.
\end{equation}
 
Before we continue, we first define the following partial ordering between two positive Hermitian operators.
\begin{defn}\label{defn:partial_ordering}
    Let $W,\widetilde{W}$ be positive Hermitian operators. We define the partial order $W \succeq \widetilde{W} \succeq \mathbf{0}$ (where $\mathbf{0}$ is the zero matrix) such that the expectation values satisfy $\Tr(\rho W) \geq \Tr(\smash[t]{\rho \widetilde{W}})\geq 0$ for any density matrix $\rho$. 
\end{defn}

We now consider the following (unnormalized) operator $\mathcal{W}$,
\begin{equation}\label{eqn:proposed_entanglement_witness_general}
    \mathcal{W} := \bigotimes_{j=1}^N W_j - \bigotimes_{j=1}^N \widetilde{W}_j,
\end{equation}
where $W_j \succeq \widetilde{W}_j$ for all $j$. That is, $\mathcal{W}$ is a difference of two product observables, which we dub the minuend and subtrahend operators, and that the local observables in each are positive semidefinite. A key observation is that $\Tr(\mathcal{W}\rho) \geq 0$ for any fully separable state $\rho$ in $\mathcal{S}$: a necessary condition for full separability. We can simply take its contraposition, giving a sufficient condition for entanglement. We now present the main proposition of this work \cite{casapaoMSc}:

\begin{prop}\label{prop:proposed_entanglement_witness}
    Let $\mathcal{W} = \bigotimes_{j=1}^N W_j - \bigotimes_{j=1}^N \widetilde{W}_j$, where $W_j \succeq \widetilde{W}_j$ for all $j$, and suppose that $\bigotimes_j W_j \not\succeq \bigotimes_j\widetilde{W}_j$. If $\Tr(\mathcal{W}\rho)< 0$, then $\rho$ is entangled.
\end{prop}

We have essentially described $\mathcal{W}$ as an EW, with $\mathcal{W}$ having at least one negative eigenvalue to be able to detect entanglement (unless we have the condition $\bigotimes_j W_j \succeq \bigotimes_j\widetilde{W}_j$, which will make $\mathcal{W}$ positive semidefinite). While both the minuend and subtrahend operators are locally decomposable and positive semidefinite, evaluating the difference of their expectation values can reveal the underlying entanglement. We can also exploit the fact that the quantities 
\begin{equation}\label{eqn:coincidence_probabilities_min_sub}
    {\Tr(\frac{\bigotimes_j W_j}{\Pi_{j}\Tr W_j} \rho)}, \qquad {\Tr(\frac{\bigotimes_j \widetilde{W}_j}{\Pi_{j}\Tr \widetilde{W}_j} \rho)}
\end{equation}
are overlaps of two valid product states, ${\bigotimes_j W_j}/{\Pi_{j}\Tr W_j}$ and ${\bigotimes_j \widetilde{W}_j}/{\Pi_{j}\Tr \widetilde{W}_j}$, with the state of interest $\rho$, and so estimating both can (in principle) be physically realizable, for example through interferometry. 

%--------------------%

\section{Witness construction}
\label{sec:witness_construction}

To construct our proposed EWs, we rephrase the partial ordering rule in Definition~\ref{defn:partial_ordering} based on our chosen density matrix parameterization \cite{casapaoMSc}. Here, we choose a generalized Bloch-vector parameterization \cite{Bertlmann2008bloch,bruning2012parametrizations}, with
\begin{equation}
    \rho = \frac{I}{d} +\frac{1}{2}\bm{\tau}\cdot \bm{\Gamma},
\end{equation}
where $\{\Gamma_k\}$ are the $d^2-1$ generalized Pauli operators satisfying $\Tr(\Gamma_k\Gamma_{k'})=2\delta_{kk'}$. We also parameterize the local operators $W,\widetilde{W}$ in a similar manner, with
\begin{align}
    W &= \nu_0 I + \bm{\nu}\cdot\bm{\Gamma},\nonumber\\
    \widetilde{W} &= \widetilde{\nu}_0 I + \bm{\widetilde{\nu}}\cdot\bm{\Gamma}.
\end{align}
The Bloch vectors $\bm{\tau},\bm{\nu},\bm{\widetilde{\nu}}$ ensure the positivity of these operators. Imposing the rule $W \succeq \widetilde{W}$ gives
\begin{equation}\label{eqn:partial_order_direct_apply}
    (\nu_0 - \widetilde{\nu}_0) + (\bm{\nu} - \bm{\widetilde{\nu}})\cdot \bm{\tau} \geq 0.
\end{equation}
Since we want Ineq.~\eqref{eqn:partial_order_direct_apply} to be true for all $\rho$, it is sufficient to consider the case $(\nu_k - \widetilde{\nu}_k)\tau_k = -\abs{\nu_k - \widetilde{\nu}_k}\abs{\tau_k}$ for all $k$. We observe that 
\begin{equation}
    0\leq \sum_{k=1}^{d^2 -1 }\frac{\abs{\nu_k-\widetilde{\nu}_k}}{\nu_0 - \widetilde{\nu}_0}\abs{\tau_k}\leq 1.
\end{equation}
We can rewrite the above inequality into
\begin{equation}\label{eqn:partial_order_direct_apply_rewrite}
    0\leq \sum_{k=1}^{d^2 - 1}\left[\frac{2(d -1)}{d}\frac{\abs{\nu_k - \widetilde{\nu}_k}}{\nu_0 - \widetilde{\nu}_0}\right]\abs{\tau_k} \leq \frac{2(d -1)}{d}.  
\end{equation}
Now, using the condition for purity, ${1}/{d}\leq \Tr(\rho^2)\leq 1$, we find that 
\begin{equation}
    0\leq \sum_{k=1}^{d^2-1}\abs{\tau_k}^2\leq \frac{2(d -1)}{d}. 
\end{equation}
We immediately see that the expression between the square brackets in Ineq.~\eqref{eqn:partial_order_direct_apply_rewrite} must have the Bloch radius $\sqrt{2(d-1)/d}$ as its upper bound. Therefore, the condition
\begin{equation}\label{eqn:partial_order_bound}
    \abs{\nu_k - \widetilde{\nu}_k} \leq (\nu_0 - \widetilde{\nu}_0)\sqrt{\frac{d}{2(d -1)}}
\end{equation}
for all $k$ satisfies the partial ordering rule.

We provide a few remarks. Constructing EWs using this parameterization inherits the problem of ensuring the positivity of the density matrices and the local operators. While positivity can be easily guaranteed for two-level systems (i.e., that a Bloch vector must be contained within a Bloch ball), explicitly showing positivity for $d\geq 3$ systems is not as straightforward. Tools such as Vieta's formulas may be useful in this regard. Other parameterizations may also be used, such as those discussed in Ref.~\cite{bruning2012parametrizations}, which opens new directions for further research. 
%There is also room to consider finding an alternative to the bound in Ineq.~\eqref{eqn:partial_order_bound}.(in particular, the Jarlskog parameterization)

By design, our proposed EWs have two operators that are already decomposed locally. This appears advantageous in experimentally detecting entanglement since we only need to measure the coincidence probabilities in \eqref{eqn:coincidence_probabilities_min_sub}, i.e., we only need two measurement settings. However, this advantage in directly implementing the EWs is truly present if the local operators are experimentally accessible themselves, and so it becomes relevant to decompose witnesses into local von Neumann measurements (LvNMs) \cite{terhal2002detecting,guhne2002detection,guhne2003investigating}. For $d=2$, a particularly favored witness decomposition that can be easily measured are the tensor products of local Pauli operators. Our chosen parameterization naturally takes these LvNMs into account.

\begin{exa}\label{exa:bell-states-witnesses}
    The Bell state $\ket*{\Phi^+} = (\ket{00}+\ket{11})/\sqrt{2}$ can be detected by EWs that satisfy the inequality $\nu_{1,0}\nu_{2,0} + \bm{\nu}_1\cdot \bm{\nu}_2 - 2\nu_{1,y}\nu_{2,y} < \widetilde{\nu}_{1,0}\widetilde{\nu}_{2,0} + \bm{\widetilde{\nu}}_1\cdot\bm{\widetilde{\nu}}_2 - 2\widetilde{\nu}_{1,y}\widetilde{\nu}_{2,y}$ following the parameterization in this section. An example is the EW 
    \begin{align}
        \mathcal{W}_{\Phi^+} =&\left(\frac{8387}{8192}I +\frac{41}{64}(\sigma_x +\sigma_z)\right)^{\otimes 2}\nonumber\\
        &-\left(I +\frac{85}{128}(\sigma_x +\sigma_z)\right)^{\otimes 2}, \label{eqn:example_witness_Phiplus}
    \end{align}
    with $\Tr(\mathcal{W}_{\Phi^+}\ketbra{\Phi^+}{\Phi^+}) \approx -0.0130$. We also observe that $\mathcal{W}_{\Phi^+}$ does not witness the Bell state $\ket*{\Psi^-} = (\ket{01}-\ket{10})/\sqrt{2}$ since $\Tr(\mathcal{W}_{\Phi^+}\ketbra{\Psi^-}{\Psi^-}) \approx 0.1093$, and therefore $\ket*{\Psi^-}$ cannot be an eigenstate that corresponds to any of the negative eigenvalues of $\mathcal{W}_{\Phi^+}$. We can also determine the robustness of $\mathcal{W}_{\Phi^+}$ to noise. Taking a white noise model $\rho(\psi)=(1-\varepsilon)\ketbra{\psi}{\psi}+\varepsilon I/4$ for some target pure state $\ket{\psi}$, we find $\varepsilon \approx 0.2123$ as the maximal tolerance of $\mathcal{W}_{\Phi^+}$ when witnessing $\ket*{\Phi^+}$. This is smaller compared to the known tolerance of $\varepsilon = 2/3$ achieved by the projector-based EW $\mathcal{W}_{\Phi^+}^{\mathrm{proj}}=I/2 - \ketbra{\Phi^+}{\Phi^+}$ (beyond which the noisy state becomes fully separable).

    On the other hand, proposed EWs that detect the Bell state $\ket*{\Psi^-}$ must satisfy the condition $\nu_{1,0}\nu_{2,0} - \bm{\nu}_1\cdot \bm{\nu}_2< \widetilde{\nu}_{1,0}\widetilde{\nu}_{2,0} - \bm{\widetilde{\nu}}_1\cdot\bm{\widetilde{\nu}}_2$. The EW
    \begin{align}
        \mathcal{W}_{\Psi^-} =&\left(\frac{607}{512}I -\frac{107}{128}(\sigma_x +\sigma_z)\right)^{\otimes 2}\nonumber\\
        &-\left(I -\frac{85}{128}(\sigma_x +\sigma_z)\right)^{\otimes 2} 
    \end{align}
    is one example, with $\Tr(\mathcal{W}_{\Psi^-}\ketbra{\Psi^-}{\Psi^-}) \approx -0.1101$. Moreover, we have $\Tr(\mathcal{W}_{\Psi^-}\ketbra{\Phi^+}{\Phi^+}) \approx 0.9211$, and therefore $\mathcal{W}_{\Psi^-}$ does not witness $\ket*{\Phi^+}$. This EW also tolerates some noise, with $\varepsilon\approx0.2135$ as its maximal noise tolerance when witnessing $\ket*{\Psi^-}$.  $\square$
\end{exa}

Notice that we only need one LvNM statistics to post-process in Example~\ref{exa:bell-states-witnesses}, namely for the LvNM $((\sigma_x+\sigma_z)/\sqrt{2})^{\otimes 2}$, to determine the expectation values of the proposed EWs. We compare this with $\mathcal{W}_{\Phi^+}^{\mathrm{proj}}$ which, although more robust to noise, requires three LvNM settings, e.g., $\sigma_j^{\otimes 2}$ for $j\in\{x,y,z\}$. From an experimental point of view, using both $\mathcal{W}_{\Phi^+}$ and $\mathcal{W}_{\Psi^-}$ may be overall cost-effective in terms of implementing the LvNM settings, in particular in situations where there is only a need to certify high-quality entanglement upon generation. For example, the EW $\mathcal{W}_{\Phi^+}$ can detect fidelities that are no less than about 0.8408 with respect to $\ket*{\Phi^+}$. 

We note that the total number of LvNMs can vary depending on our constructed EWs. Given the structure of our proposed set of EWs, it is also possible to construct an EW that requires just as many number of LvNMs (without doing any optimization) as quantum tomography. Hence, to truly claim advantage over tomography, we must also take into account that the EWs constructed require as little number of LvNMs as possible, just as the above examples.

%--------------------%

\section{Genuine multipartite entanglement}
\label{sec:genuine_entanglement}

It is important to point out that Proposition~\ref{prop:proposed_entanglement_witness} does not guarantee the structure of the entanglement witnessed. Unlike in bipartite systems, the entanglement structure of a multipartite system can be much richer since it can be entangled on some of its subsystems and is therefore only partially separable, or is not partially separable anywhere and is therefore fully genuinely entangled \cite{horodecki2009quantum}. Suitable EWs for genuine multipartite entanglement need to ensure that a state cannot be biseparable across all of its possible bipartitions, which is much more computationally involved than what is required to simply detect entanglement. EW constructions have been proposed to systematically detect genuine entanglement while remaining efficient over state tomography \cite{bourennane2004experimental,huber2014witnessing}.

In this work, we present a strategy to witness genuine multipartite entanglement using our proposed EWs. We begin with a necessary criterion on biseparability.

\begin{lem}\label{lem:biseparability_proposed_witness_general}
    If $\rho$ is biseparable, then there exists a distinct bipartition $A|\overline{A}$ of the Hilbert space $\mathcal{H} = \mathcal{H}_{A}\otimes \mathcal{H}_{\overline{A}}$ such that for any EW $\mathcal{W}_{A|\overline{A}} := W_{A}\otimes W_{\overline{A}} - \widetilde{W}_{A}\otimes \widetilde{W}_{\overline{A}}$ we have $\Tr(\rho\smash{\mathcal{W}_{A|\overline{A}}})\geq 0$, where $W_{A}\succeq \widetilde{W}_{A} \succeq \mathbf{0}$ and $W_{\overline{A}}\succeq \widetilde{W}_{\overline{A}}\succeq \mathbf{0}$.
\end{lem}

\noindent We can easily see this by writing $\rho$ as a convex sum of $(A,\overline{A})$-tensor products,
\begin{equation}
    \rho = \sum_q\lambda_q\rho_{A}^{(q)}\otimes\rho_{\overline{A}}^{(q)},
\end{equation}
and using the partial ordering requirement upon evaluating $\Tr(\rho\smash{\mathcal{W}_{A|\overline{A}}})$. Taking the contrapositive of the above lemma gives a sufficient condition for genuine entanglement. However, this genuine entanglement criterion requires probing all $2^{N-1}-1$ possible bipartitions $A|\overline{A}$, which is computationally expensive. 

We can further simplify the criterion by considering two local subsystems $\mathcal{H}_k$ and $\mathcal{H}_{k'}$ separated by a bipartition, i.e., $k\in A$ and $k'\in \overline{A}$. With $W_k\succeq\widetilde{W}_{k}$ and $W_{k'}\succeq\widetilde{W}_{k'}$, we can define a new EW $\mathcal{W}^{(k,k')} := (W_k\otimes W_{k'} - \widetilde{W}_k\otimes \widetilde{W}_{k'}) \otimes \mathbf{I}$, where $\mathbf{I}$ indicates a padding of identity matrices. We have the following observation.

\begin{obs}\label{obs:biseparable-pairwise-check}
    If $\rho$ is biseparable, then there exist distinct $\mathcal{H}_k$ and $\mathcal{H}_{k'}$ separated by a bipartition of $\rho$ such that for any EW $\mathcal{W}^{(k,k')} := (W_k\otimes W_{k'} - \widetilde{W}_k\otimes \widetilde{W}_{k'}) \otimes \mathbf{I}$ defined above we have $\Tr\,(\mathcal{W}^{(k,k')}\rho)\geq 0$.
\end{obs}

\noindent Taking the contrapositive of Observation~\ref{obs:biseparable-pairwise-check} gives a new criterion for genuine entanglement:

\begin{obs}\label{obs:genuine-ent-pairwise-check}
    For every pair $(k,k')$ take a corresponding EW $\mathcal{W}^{(k,k')}$ defined earlier. If $\Tr\,(\mathcal{W}^{(k,k')}\rho)< 0$ for all $(k,k')$, then $\rho$ is genuinely multipartite entangled.
\end{obs}

Observation~\ref{obs:genuine-ent-pairwise-check} simply scans for pairwise entanglement over all possible pairs of subsystems. This strategy reduces the number of checks needed to reveal genuine multipartite entanglement from an exponential to a quadratic scaling, that is, to just $\binom{N}{2}$ of such $(k,k')$ pairs. 

Obviously, not all genuine entanglement can be detected following this criterion. One notable caveat is that Observation~\ref{obs:genuine-ent-pairwise-check} cannot resolve genuine multipartite entanglement of states whose reductions in the $(k,k')$-composite system are separable, even though it is possible to detect entanglement via Proposition~\ref{prop:proposed_entanglement_witness}. An example is the $N$-qubit GHZ state $\ket{\mathrm{GHZ}_N} = (\ket{00\dots 0}+\ket{11\dots 1})/\sqrt{2}$, which has an $N$-way entanglement but no pairwise entanglement. Genuine entanglement of such states may be verified in some other way, such as through the purity-based criterion in Ref.~\cite{flores2016mixtures}, where they considered entangled states that become separable after tracing out a single subsystem. On the other hand, it is possible for the above observation to guarantee genuine entanglement of states such as the $N$-qubit W state, in which the entanglement remains robust even after tracing out $N-2$ of the subsystems. Worth noting is that our criterion can detect genuine entanglement of states that are outside the scope of Ref.~\cite{flores2016mixtures}. 

\begin{exa} We consider the three-qubit W state $\ket{\mathrm{W}_3}=(\ket{001}+\ket{010}+\ket{100})/\sqrt{3}$. A set of EWs that detects this state is
    \begin{align}
        \mathcal{W}_{\mathrm{W}_3}^{(k,k')} =&\left(\left(\frac{526849}{524288}I +\frac{170497}{262144}\sigma_x -\frac{773}{1024}\sigma_z\right)^{\otimes 2}\right.\nonumber\\
        &\left.-\left(I +\frac{671}{1024}\sigma_x -\frac{3}{4}\sigma_z\right)^{\otimes 2} \right)_{(k,k')}\otimes I,
    \end{align}
    with $\Tr(\mathcal{W}_{\mathrm{W}_3}^{(k,k')}\ketbra{\mathrm{W}_3}{\mathrm{W}_3}) \approx -0.0026$ for any $(k,k')$. Moreover, this EW set also ensures that $\ket{\mathrm{W}_3}$ is genuinely multipartite entangled via Observation~\ref{obs:genuine-ent-pairwise-check}. 
    
    We can also test for robustness to noise. Taking a white noise model $\rho(\varepsilon)=(1-\varepsilon)\ketbra{\mathrm{W}_3}{\mathrm{W}_3}+\varepsilon I/8$, we find $\varepsilon\approx 0.2108$ as the maximal tolerance of all these EWs when witnessing $\ket{\mathrm{W}_3}$. On the other hand, the projector-based EW $\mathcal{W}_{\mathrm{W}_3}^{\mathrm{proj}}=(2/3)I - \ketbra{\mathrm{W}_3}{\mathrm{W}_3}$ detects genuine entanglement up to $\varepsilon=8/21\approx 0.3810$. 
    
    While $\mathcal{W}_{\mathrm{W}_3}^{\mathrm{proj}}$ optimally requires five LvNMs \cite{bourennane2004experimental}, only four LvNMs (for example, $\sigma_z^{\otimes 3}$, $\sigma_x\otimes\sigma_x\otimes\sigma_z$, $\sigma_x\otimes\sigma_z\otimes\sigma_x$, and $\sigma_z\otimes\sigma_x\otimes\sigma_x$) are needed to account for all the three EWs $\{\mathcal{W}_{\mathrm{W}_3}^{(k,k')}\}$. However, it is also possible to construct an EW for $\ket{\mathrm{W}_3}$ that can perform better in terms of both the noise tolerance and the number of LvNMs: for example an EW constructed following the stabilizer formalism \cite{toth2005entanglement}. $\square$
\end{exa}

%--------------------%

\section{Witness improvements}
\label{sec:witness_improvements}

We now discuss a simple method to linearly improve our proposed EWs while retaining their minuend-subtrahend form \cite{casapaoMSc}. An EW $\mathcal{W}_f$ is defined as finer than another EW $\mathcal{W}_c$ if and only if $\mathcal{W}_f$ detects all states detected by $\mathcal{W}_c$. Following Lemma 2 of Ref.~\cite{lewenstein2000optimization}, it was shown that $\mathcal{W}_f$ is finer than EW $\mathcal{W}_c$ if and only if there exist some positive operator $P$ and real constant $\varepsilon$, where $0\leq\varepsilon<1$, such that 
\begin{equation}
    \mathcal{W}_c = (1-\varepsilon)\mathcal{W}_f + \varepsilon P.
\end{equation}
This means that $\Tr(\mathcal{W}_f\rho) \leq \Tr(\mathcal{W}_c\rho)< 0$ for all entangled $\rho$ witnessed by $\mathcal{W}_c$ since $\Tr(P\rho) \geq 0$.

Consider a bipartite system $\mathcal{H}_1\otimes\mathcal{H}_2$, and let $\mathcal{W}_c = (W_{1}^c\otimes W_{2}^c -\widetilde{W}_{1}^c\otimes\widetilde{W}_{2}^c)/n_{c}$ be our proposed witness, with $n_c$ as a normalization constant. Suppose further that the finer EW $\mathcal{W}_f$ preserves our proposed minuend-subtrahend structure, i.e., $\mathcal{W}_f = (W_{1}^f\otimes W_{2}^f -\widetilde{W}_{1}^f\otimes\widetilde{W}_{2}^f)/n_{f}$. Then, 
\begin{align}
    &W_{1}^c\otimes W_{2}^c -\widetilde{W}_{1}^c\otimes\widetilde{W}_{2}^c\nonumber\\
    &= (1-\varepsilon)\frac{n_c}{n_f}(W_{1}^f\otimes W_{2}^f -\widetilde{W}_{1}^f\otimes\widetilde{W}_{2}^f) + \varepsilon n_c P.
\end{align}
We can take a simple case in which
\begin{align}
    W_{1}^f\otimes W_{2}^f &:= (1-\zeta)W_{1}^c\otimes W_{2}^c, \nonumber\\
    \widetilde{W}_{1}^f\otimes\widetilde{W}_{2}^f &:= (1+\xi)\widetilde{W}_{1}^c\otimes\widetilde{W}_{2}^c,
\end{align}
with the real constants $\zeta,\xi$ to be determined later. A choice for $\varepsilon$ is $\varepsilon = 1 - (n_f/n_c)$, where we have to further impose the condition $n_f \leq n_c$ to satisfy $0\leq \varepsilon < 1$. We immediately see that 
\begin{equation}
    P = \frac{{\zeta} W_{1}^c\otimes W_{2}^c + {\xi}\widetilde{W}_{1}^c\otimes\widetilde{W}_{2}^c}{\zeta \Tr(W_{1}^c\otimes W_{2}^c)+\xi\Tr(\widetilde{W}_{1}^c\otimes \widetilde{W}_{2}^c)}
\end{equation}
is our desired operator. Here, we already evaluated $n_c-n_f$, giving us the denominator in the above expression. Finally, we restrict $\zeta,\xi$ for $P\succeq \mathbf{0}$ to be true. It is obvious that $\zeta,\xi$ cannot be both negative. On the other hand, $P\succeq \mathbf{0}$ is guaranteed whenever both $\zeta,\xi$ are non-negative. For the latter case, we can consider 
\begin{equation}\label{eqn:condition-witness-improvement1}
    (1-\zeta)W_{1}^c\succeq (1+\xi)\widetilde{W}_{1}^c,
\end{equation}
in which we further observe that $0\leq \zeta <1$ must also be satisfied. All $(\zeta,\xi)$ values that satisfy Condition~\eqref{eqn:condition-witness-improvement1} correspond to valid witnesses that have a larger set of detectable states than $\mathcal{W}_c$. We can then iterate the same process for $W_{2}^c$ and $\widetilde{W}_{2}^c$. This observation can also be easily extended to the general $N$-partite case. We remark that $\mathcal{W}_f(\zeta,\xi)$ may not be an optimal EW as defined in Ref.~\cite{lewenstein2000optimization}, since optimality requires that no other finer EW can be found.

We can also improve our EWs nonlinearly. We revisit the results in Ref.~\cite{guhne2006nonlinear}, in which nonlinearities are introduced to the EW inequalities to detect a larger set of entangled states. Suppose that our proposed EW $\mathcal{W}=(W_1\otimes W_2 - \widetilde{W}_1\otimes \widetilde{W}_2)/n_{\mathcal{W}}$ can detect the state $\ket{\phi}$. One such nonlinear functional is
\begin{equation}\label{eqn:nonlinear_functional_witness}
    \mathcal{F}^{(1)}(\rho) := \Tr(\mathcal{W}\rho)- \frac{1}{s(\psi)}\Tr(X\rho)\Tr(X^{\dagger}\rho)
\end{equation}
for some chosen pure state $\ket{\psi}\in \mathcal{H}_1\otimes \mathcal{H}_2$, with $\mathcal{H}_1\cong \mathcal{H}_2$, and $s(\psi)$ is the square of the pure state's largest Schmidt coefficient. Here, 
\begin{equation}\label{eqn:nonlinear-hermitian-term}
    X := (I\otimes \Lambda )(\ketbra*{\psi}{\phi}),
\end{equation}
where $\Lambda$ is the Choi-Jamio\l{}kowski map to $\mathcal{W}$, given by
\begin{equation}
    \Lambda(\rho) = d\Tr_1\left(\mathcal{W}(\rho^{T}\otimes I)\right).
\end{equation}
The corresponding map to $\mathcal{W}$ is
\begin{equation}
    \Lambda(\rho) = \frac{d}{n_{\mathcal{W}}}\left[\Tr(W_{1}\rho^T)W_2 - \Tr(\smash[t]{\widetilde{W}_{1}}\rho^T)\widetilde{W}_2\right].
\end{equation}
An interesting feature of this map is that it trivially decomposes into a difference of two completely positive (CP) maps (in fact, it is known that any positive map can be written as a difference of two CP maps \cite{chruscinski2007structure}).

\begin{exa}
    The generalized Bell states can be written as
    \begin{equation}
        \ket{\Phi_{m,n}} := (I\otimes \Omega_{m,n})\ket*{\Phi_d^+}
    \end{equation}
    where $\Omega_{m,n} := \sum_k \exp(2\pi i kn/d)\ketbra{k}{(k+m)\mod d}$ belongs to the set of $d^2$ Heisenberg-Weyl operators, and $\ket*{\Phi_d^+} = (1/\sqrt{d})\sum_k \ket{k}\otimes\ket{k}$ is the $d$-maximally entangled state. Suppose that some EW $\mathcal{W}$ has the proposed minuend-subtrahend form and detects $\ket*{\Phi_d^+}$. With respect to a chosen $\ket*{\Phi_{m,n}}$, and according to Eq.~\eqref{eqn:nonlinear-hermitian-term}, we get
    \begin{equation}
        X = \left(\Omega_{m,n}^{T}\otimes I\right)\mathcal{W}.
    \end{equation}    
    We then construct the nonlinear functional in Eq.~\eqref{eqn:nonlinear_functional_witness}, with $s(\psi)=1/d$. Here, it is worth noting that $X$ is also a difference of two product observables. 

    We can take the EW $\mathcal{W}_{\Phi^+}$ in Eq.~\eqref{eqn:example_witness_Phiplus} as an example. We can linearly improve $\mathcal{W}_{\Phi^+}$ to a finer $\mathcal{W}_{\Phi^+}^f(\zeta,\xi)$ by taking $\zeta = 1/9090$ and $\xi = 1/9088$. Now, suppose that we wish to construct $\mathcal{F}^{(1)}$ based on $\ket{\Psi^+}= (\ket{01}+\ket{10})/\sqrt{2}$, so that $X= (\sigma_x \otimes I)\mathcal{W}_{\Phi^+}$ and $s(\Psi^+) = 1/2$. If we consider the convex set of Bell-diagonal states, we see an increase of about $9\%$ relative to the set of states detected by $\mathcal{W}_{\Phi^+}$. On the other hand, using $\mathcal{F}^{(1)}$ increases the detectable set by about $10\%$ relative to the set detected by $\mathcal{W}_{\Phi^+}^f$. $\square$
\end{exa} 

%--------------------%

\section{Conclusion}
\label{sec:conclusions}

We investigated a class of EWs, where each EW is formulated as a difference of two observables. These observables are decomposable into positive semidefinite local operators that obey a partial ordering rule defined over all their possible expectation values. We are motivated by the need of structurally simple EWs which can be prescriptively constructed, while in principle reducing the number of local measurements (for our case, limiting to two). We provided a simple framework to construct the proposed EWs, as well as discussed methods to improve the EWs in a linear or nonlinear fashion. We showed that these EWs can also be used to reveal genuine multipartite entanglement.

There is still room to extensively study the predictive power of our proposed EWs, in particular identifying the classes of states in which the EWs can detect. On the experimental side, directly implementing the EWs will be far from an idealized scenario, where imperfect measurement settings can potentially undermine the estimation of the EWs \cite{Rosset2012imperfect,Morelli2021Entanglement,qiu2024protecting}. Thus, it is also worth investigating the sensitivity of our proposed EWs in the presence of these systematic errors.

%--------------------%

\bibliography{witness-letter.bib}

%--------------------%

\end{document}